\documentclass{ws-p9-75x6-50}

\def\apj#1{{\em Astrophys. J.} {\bf #1}}

\def\mn#1{{\em Mon. Not. R. astr. Soc.} {\bf #1}}
\def\aa#1{{\em Astron. Astrophys.} {\bf #1}}
\def\nat#1{{\em Nature} {\bf #1}}

\def\etal{{\it et al.\/}\ }
\def\Mpc{$h^{-1}$~{\rm  Mpc}}
\def\hmpc{$h$~{\rm  Mpc$^{-1}$}}

\begin{document}

\title{The structure of the Universe on 100 Mpc Scales}

\author{Jaan Einasto} 
\address{Tartu Observatory, EE-61602, Estonia}

\maketitle

\abstracts{Observational evidence for the presence of a preferred
scale around 100~Mpc in the distribution of high-density regions in
the Universe is summarised.  Toy models with various degrees of
regularity are analysed to understand better the observational data.
Predictions of models which produce a feature on the initial power
spectrum of matter are analysed and compared with new observational
data on the CMB angular spectrum, cluster mass function, and galaxy
and cluster power spectrum.  }

\section{Introduction}

Until the mid-1970s the astronomical community accepted the paradigm
that galaxies are distributed in space more or less randomly and that
about $5 - 10$~\% of all galaxies are located in clusters which are
also randomly distributed. A paradigm change occurred during a
Symposium on Large-Scale Structure of the Universe in 1977 in Tallinn.
Several teams reported results of studies of the 3-dimensional
distribution of galaxies in superclusters, and demonstrated that
superclusters consist of chains (filaments) of galaxies, groups and
clusters, and that the space between galaxy systems is devoid of any
visible galaxies (J\~oeveer \& Einasto 1978, Tarenghi \etal 1978,
Tifft \& Gregory 1978, Tully \& Fisher 1978).

The discovery of the filamentary character of superclusters and the
presence of large empty voids made obsolete the early theories of structure
formation -- the hierarchical clustering model by Peebles (1980), and
the whirl theory by Ozernoy (1978). The first impression was that the
winning theory was the pancake scenario by Zeldovich (1978).  However,
a more detailed analysis has shown that the original pancake scenario
also has weak points -- the structure forms too late and has no fine
structure of faint galaxy filaments in large voids (Zeldovich, Einasto
\& Shandarin 1982).  Then the Cold Dark Matter model was suggested and
generally accepted (Blumenthal et al 1984). Recently it was replaced
by a CDM model dominated by a cosmological term (vacuum energy) -- the
LCDM model.  However, this model has also at least one weak point: in
LCDM model superclusters are randomly distributed, whereas
observational evidence suggests that rich superclusters and voids form
a quasi-regular network of scale $\sim 100 - 130$~\Mpc\ (Einasto \etal
1997a, 1997d). Here we use the dimensionless Hubble constant $h$
defined as $H_0 = 100~h$ km~s$^{-1}$~Mpc$^{-1}$.  These data raise the
question: Is the regularity of the distribution of superclusters a
challenge for CDM models?  My lecture tries to answer this question.
First I shall summarise the observational evidence for the presence of
the 130~\Mpc\ scale, thereafter I shall discuss attempts to explain
the presence of this scale theoretically.

\section{Evidence for the 100 -- 130 Mpc Scale}

The evidence for the presence of a scale in the distribution of
superclusters has accumulated slowly.  Zeldovich, Einasto \& Shandarin
(1982) noticed that voids between superclusters have mean diameters
about 100~\Mpc.  Kopylov et al. (1988) calculated the cluster
correlation function and found a secondary peak at $\sim 125$~\Mpc,
which corresponds to superclusters on opposite sides of large voids.
Using a slightly different method Mo et al. (1992) confirmed the
presence of a feature in the cluster correlation function at $\sim
125$~\Mpc.

The most dramatic demonstration of the scale was made by Broadhurst,
\etal (1990) who measured redshifts of distant galaxies in a pencil
beam directed toward Northern and Southern galactic poles.  Their data
show that high-density regions in this direction alternate with
low-density regions with surprising regularity of a period $\sim
128$~\Mpc.  Finally, Einasto et al (1997a, 1997b, 1997d) calculated
the cluster correlation function and power spectrum of a 3-dimensional
cluster sample and confirmed the presence of a scale of $\sim
120$~\Mpc\ in the distribution of rich clusters of galaxies.

A similar phenomenon is observed in the distribution of Lyman-break
galaxies (Broadhurst \& Jaffe 2000) at high redshift, $z \approx
3$. The correlation function of these distant galaxies has secondary
peaks.  A regularity on a similar scale has been found in the
distribution of quasars by Roukema \& Mamon (2000).  All these data
suggest that this scale is primordial and co-moves with the expansion;
it can be used as a standard rod.

\section{Power Spectrum and Correlation Function of Toy Models}

The fluctuating density field can be described by the power spectrum
and the correlation function.  The power spectrum $P(k)$ characterises
the amplitude of density fluctuations for various wavenumbers
$k=2\pi/l$.  It is determined by the power indices on large and small
scales, $n_l$, $n_s$, by the transition scale $l_0$ (or wavenumber
$k_0=2\pi/l_0$) and by the amplitude $\sigma_8$ on scale 8~\Mpc.  The
correlation function $\xi(r)$ characterises the clustering of galaxies
or clusters in space; it is defined as the excess probability to find
galaxies or clusters over a random distribution of points.

\begin{figure}[ht]
\vspace*{15.5cm}
\caption{The distribution of  clusters in Voronoi and regular rod models
(upper left and right panels, respectively), respective correlation
functions (middle panels) and power spectra (lower
panels). In the lower panels a wavenumber $k=1$ corresponds to the
size of the box in the upper panels.}
\includegraphics{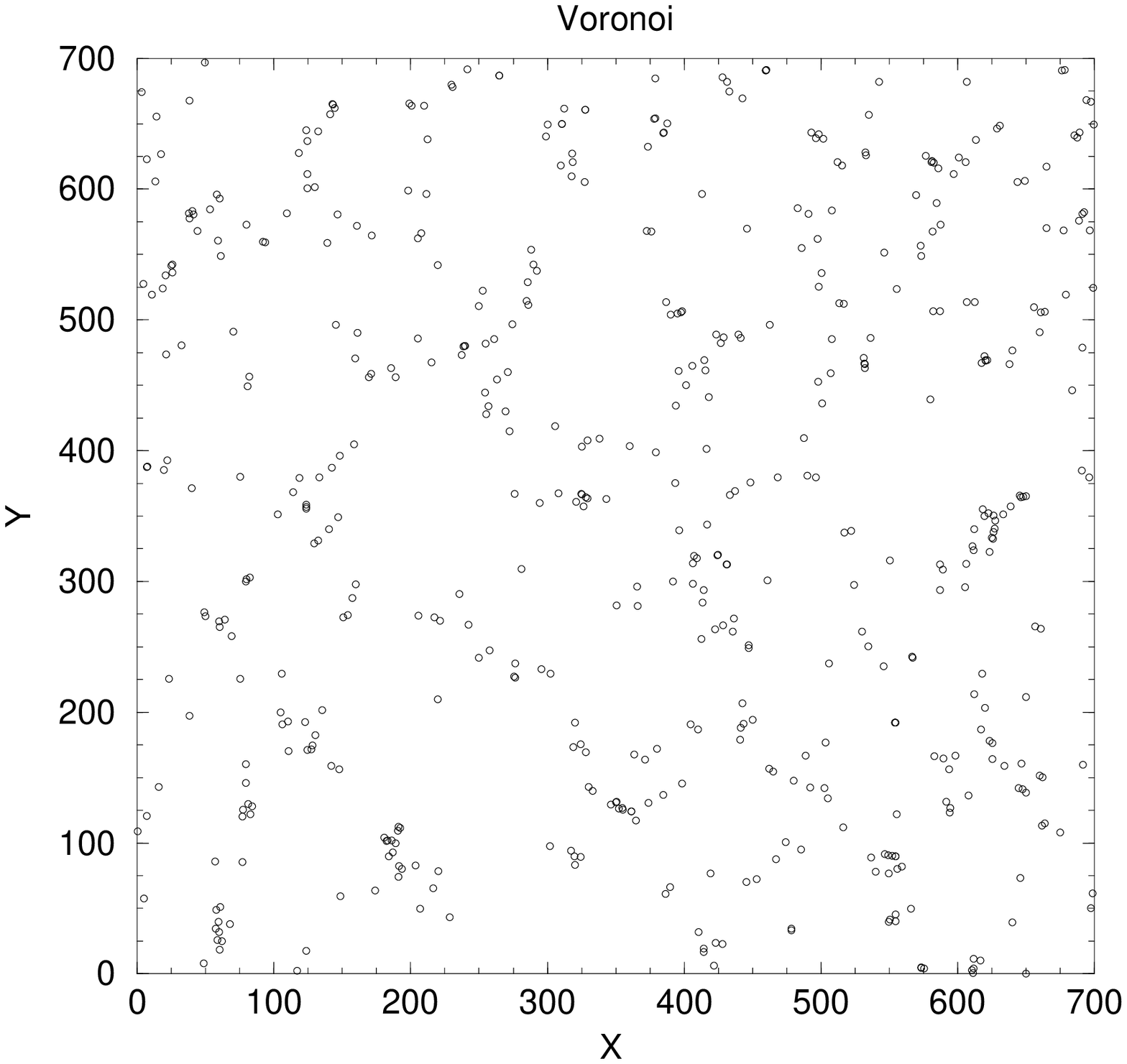}
\includegraphics{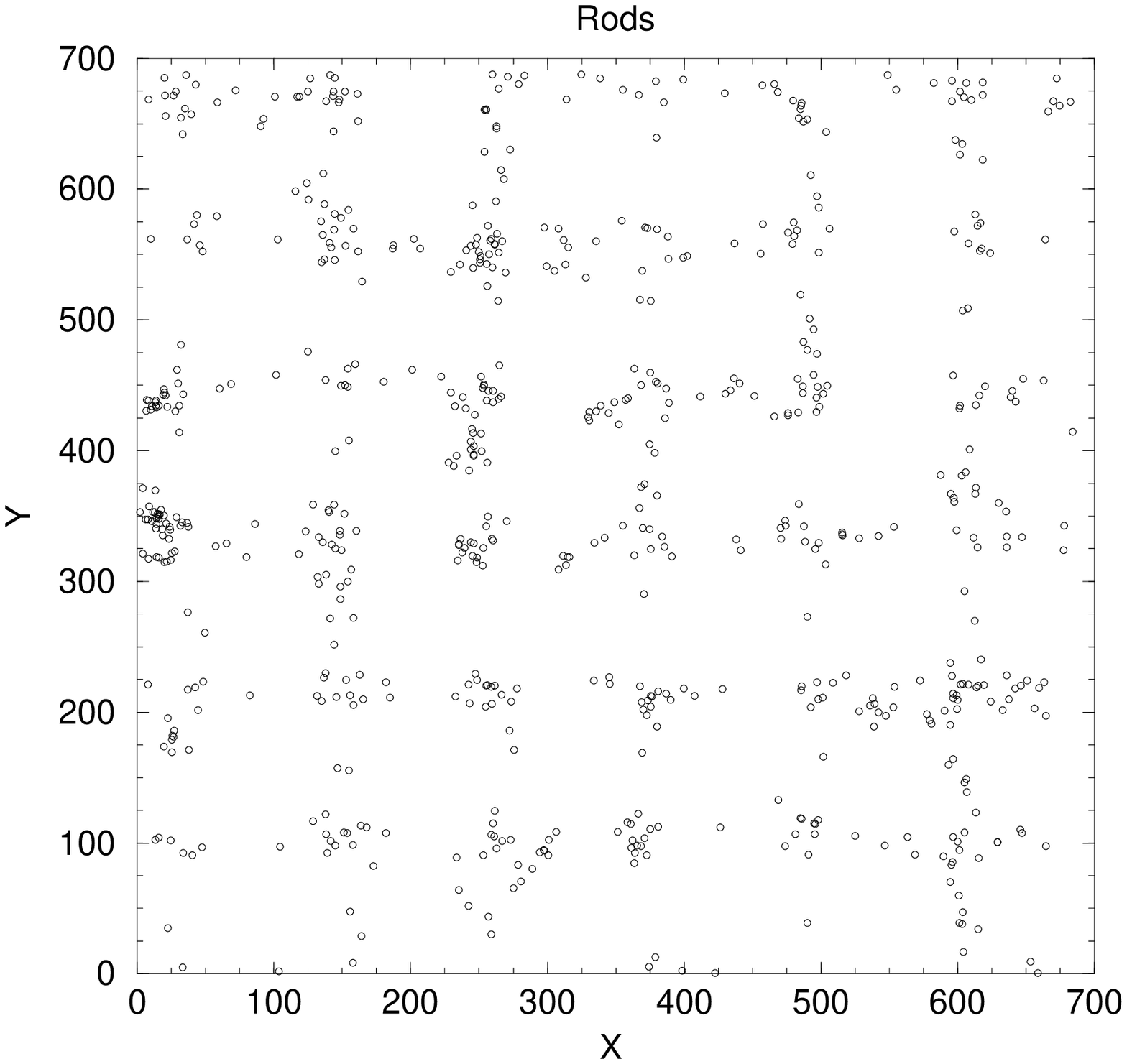}
\includegraphics{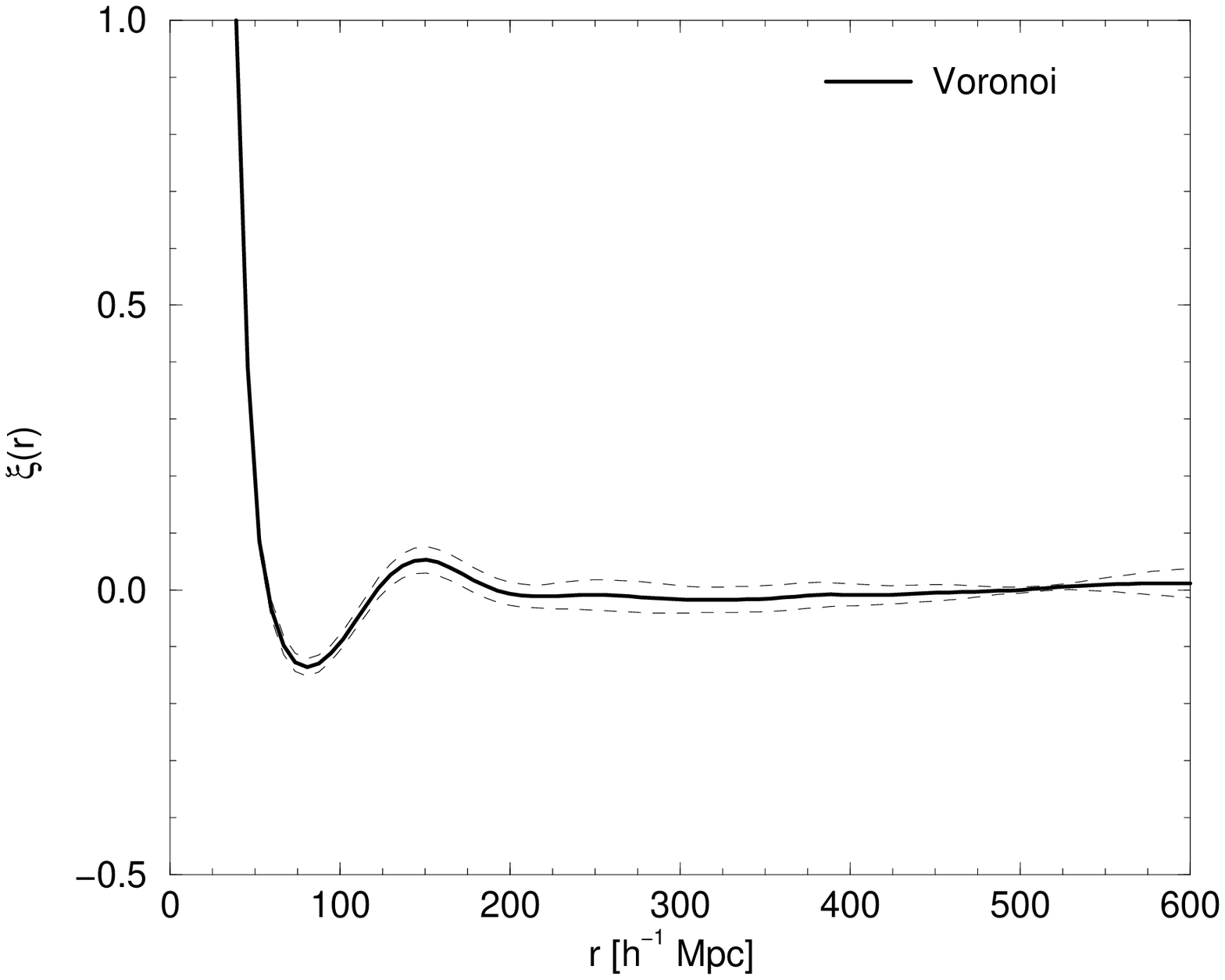}
\includegraphics{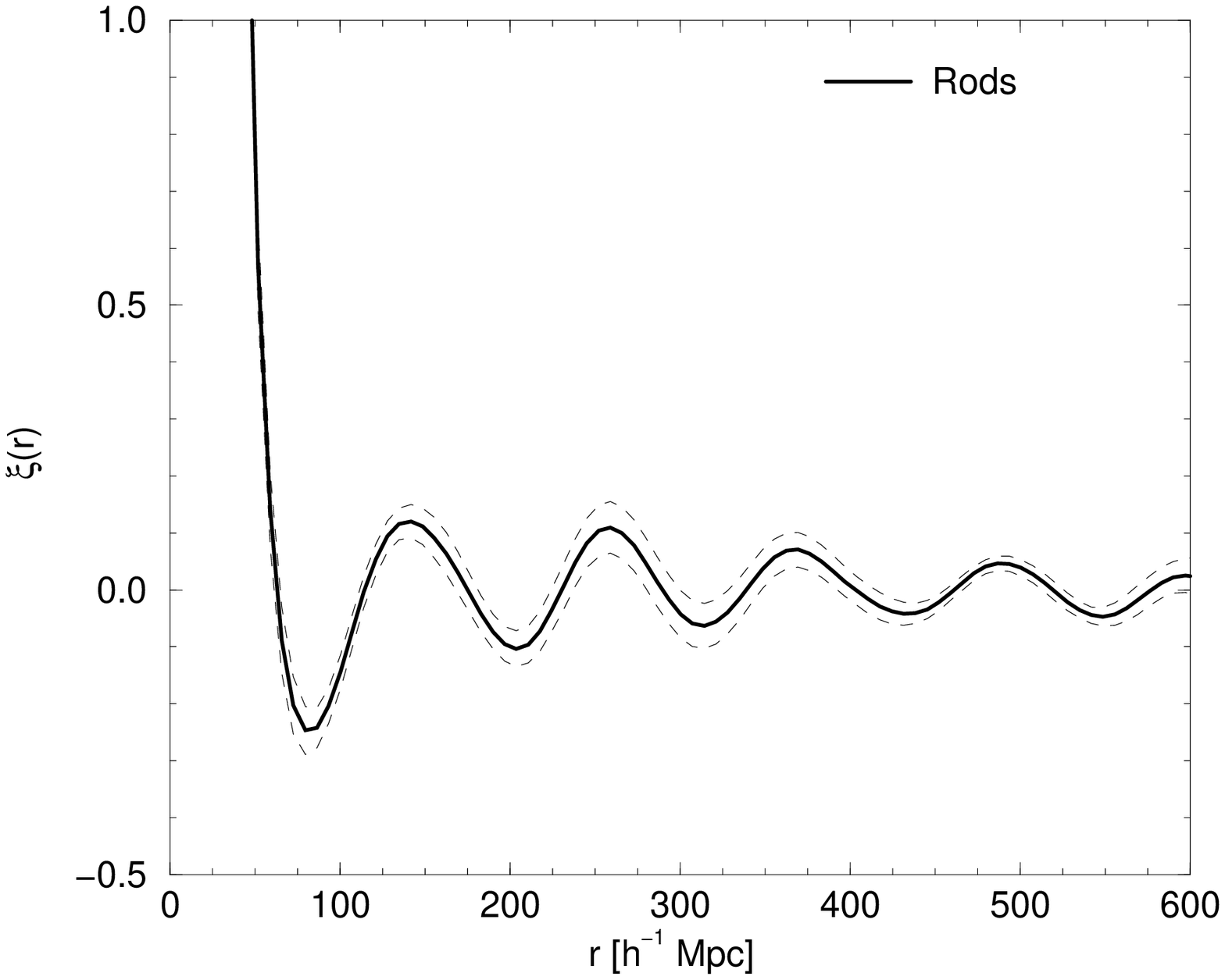}
\includegraphics{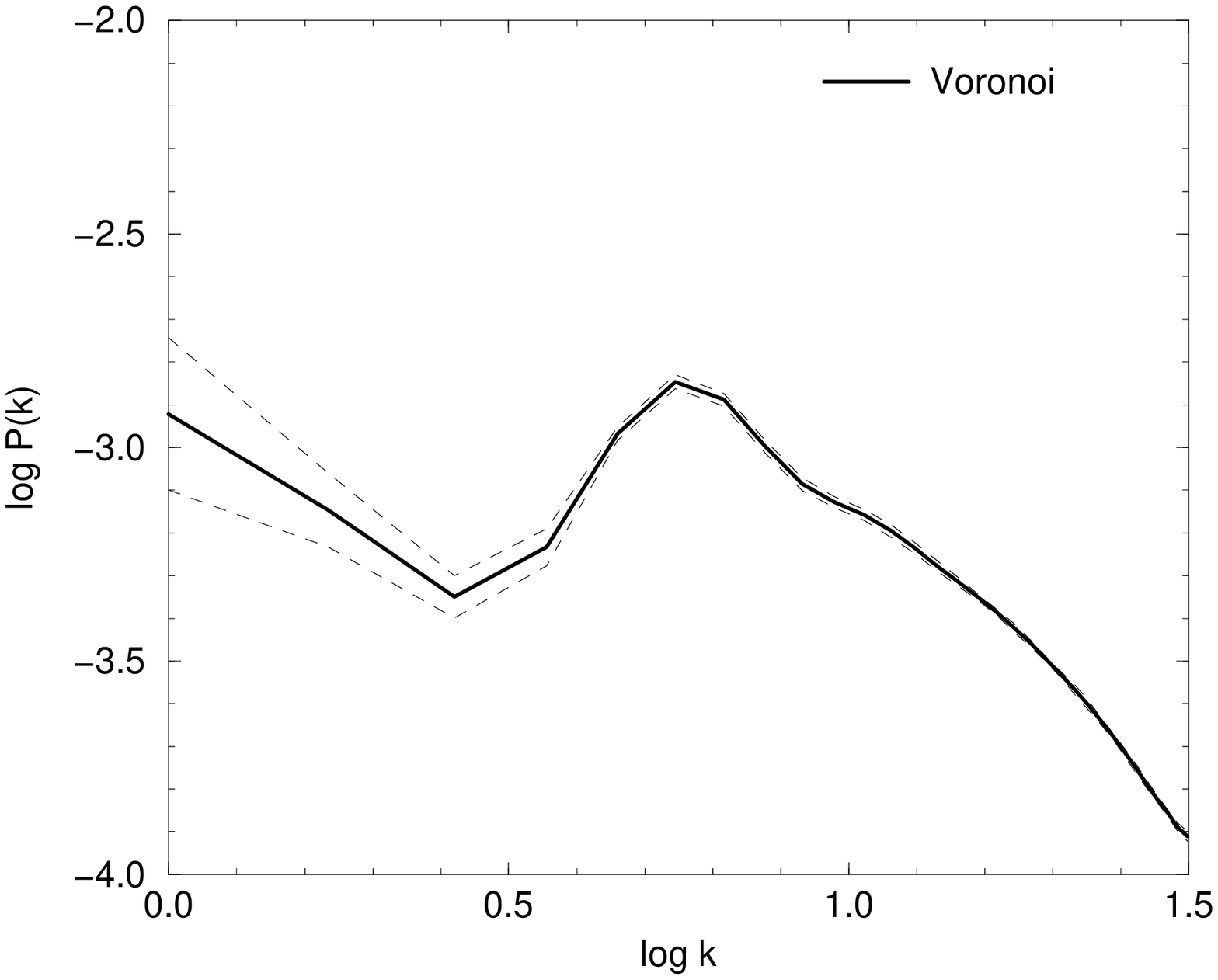}
\includegraphics{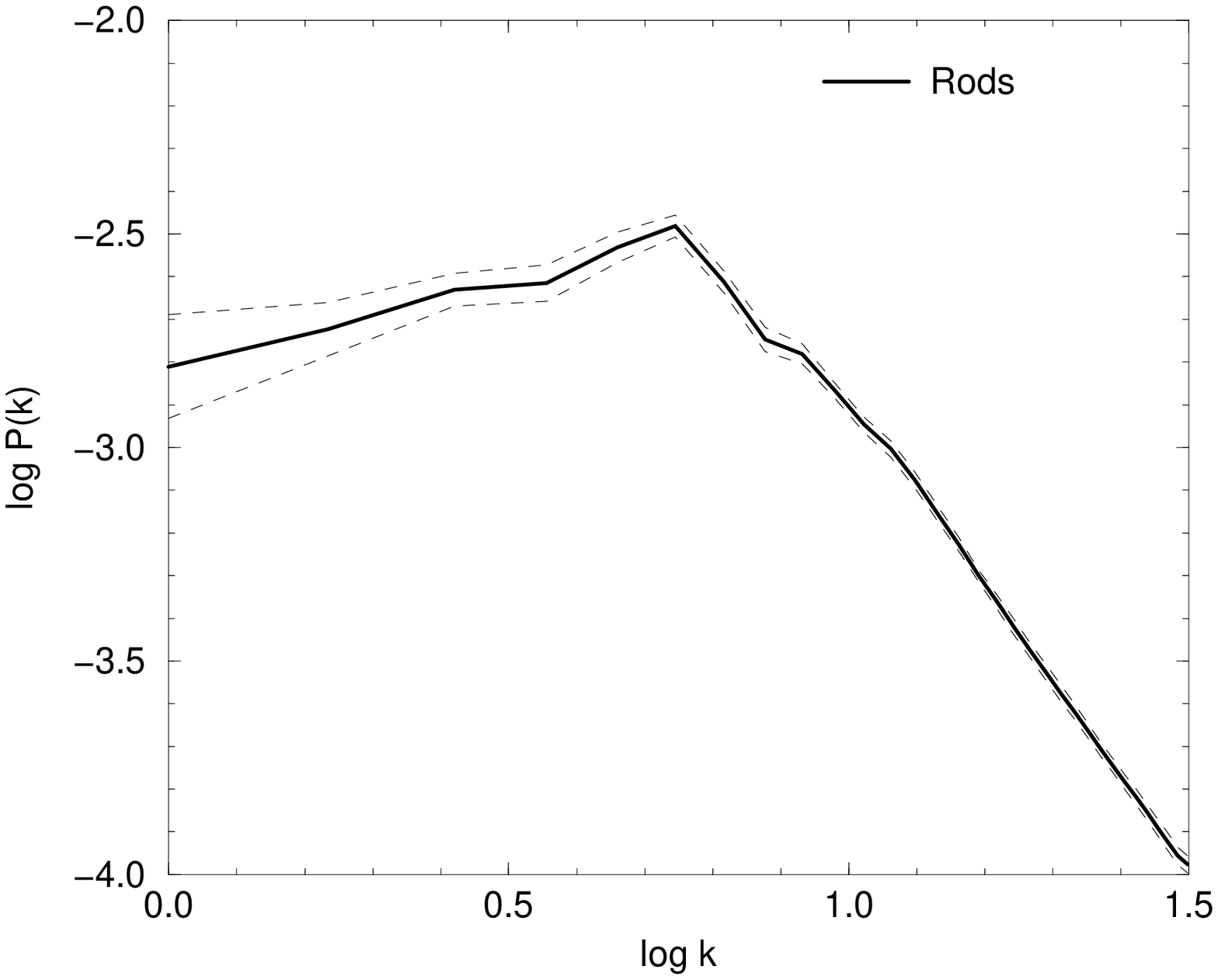}
\label{fig:corr2}
\end{figure}

In the ideal case of data not being distorted by errors and being
available over the whole space, the power spectrum and the correlation
function form mutual pairs of Fourier transformations.  In reality
they complement each other.  We are interested here in the study of
the regularity of the structure.  To understand better observational
data we shall study first several toy models of different geometry.
These toy models are: random superclusters, Voronoi tessellation
(randomly located void centers, and clusters located between voids as
far from void centers as possible), and ``regular rod models''
(superclusters located randomly along regularly spaced parallel rods
oriented along all 3 dimensions).  Details of these models were
discussed by Einasto et al. (1997c).  In all three models clusters
within superclusters are generated in an identical way; they are
distributed randomly according to an isothermal density law. Thus, on
small scales correlation functions and power spectra of all three
models are identical.  Only on larger scales the models differ.  The
random supercluster model has no built-in scale and no regularity in
the distribution of superclusters.  The Voronoi tessellation model has
a built-in scale -- the mean size of voids, but no regularity in the
distribution of superclusters as voids are distributed randomly.  The
regular rod model has a scale: the spacing of the grid of rods, and
also a regularity: the rods form a regular lattice.  In
Figure~\ref{fig:corr2} we show the distribution of clusters in a sheet
of the Voronoi and regular rod models (upper panels), the
corresponding correlation functions (middle panels) and power spectra
(lower panels).

The analysis of these toy models allows us to make the following
conclusions.  On small scales ($r<50$~\Mpc) the correlation function
and power spectrum characterise the distribution of clusters within
superclusters, while on larger scales they describe the distribution
of superclusters themselves.  If superclusters are distributed
randomly, then on large scales ($r > 50$~\Mpc) the correlation
function is almost zero, $\xi(r) = 0$, and the power spectrum $P(k)$
is flat.  The correlation function of the Voronoi model has one
secondary minimum and maximum and is zero thereafter, and the power
spectrum $P(k)$ has a sharp maximum. The correlation function of the
regular rod model is oscillating: it has regularly spaced minima and
maxima; the power spectrum has a well-defined sharp maximum on a scale
equal to the period of oscillation of the correlation function; it can
be approximated by two power law. The position of the maximum of
$P(k)$ and the separation of maxima of $\xi(r)$ correspond to the mean
diameter of voids and to the separation of superclusters across voids,
respectively. 

For comparison we can say that all CDM-type models have a smooth power
spectrum and a correlation function rather similar to those of the
random supercluster model.  Here the conflict between observations and
model is well seen.

\begin{figure*}[ht]
\vspace*{7.8cm} 
\caption{The distribution of Abell clusters (open circles) and X-ray
clusters (squares) in supergalactic coordinates.  Filled squares
denote members of rich superclusters and open squares -- isolated
X-ray clusters and members of poor systems. In order to minimize the
projection effects we plot only Abell clusters from very rich
superclusters with at least 8 members.}
\includegraphics{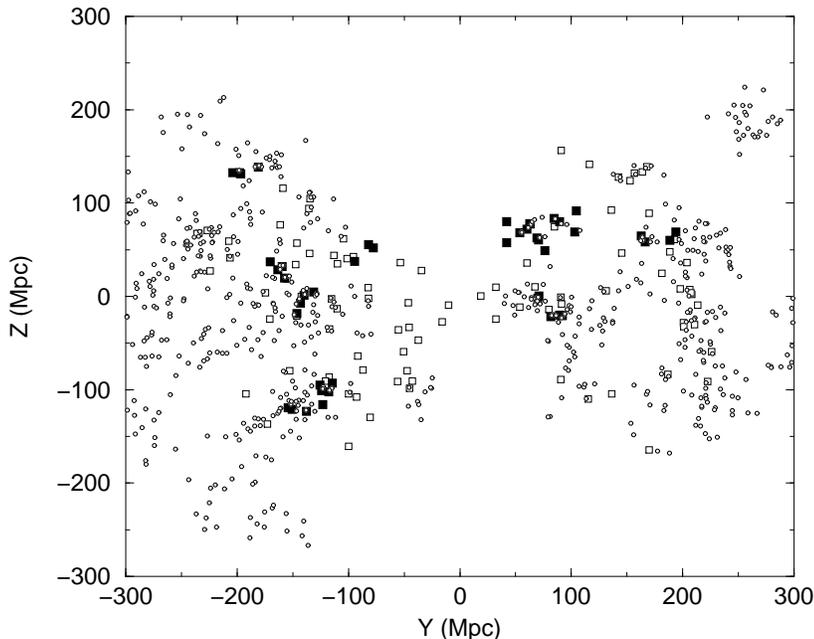}
\label{clus}
\end{figure*}

\section{Distribution of Superclusters}

\subsection{Abell Clusters}

We shall use Abell clusters of galaxies as indicators of the
distribution of high-density regions in the Universe.  The
distribution of Abell clusters in rich superclusters is shown in
Figure~\ref{clus} (open circles).  We use clusters within a distance
limit 350~\Mpc; the Northern galactic hemisphere is at right, the
Southern one at left.

\begin{figure}[ht]
\vspace*{5.0cm}
\caption{Left: power spectra of galaxies and clusters of galaxies
normalized to the amplitude of the 2-D APM galaxy power spectrum 
(references are given by Einasto \etal 1999a).  For
clarity error bars are not indicated and spectra are shown as smooth
curves rather than discrete data points.  Bold lines show spectra for
cluster data. Points with error bars show the spectrum of Abell
clusters by Miller \& Batuski (2000) adjusted to the galaxy
spectrum amplitude by a relative bias factor $b=3.2$.  Right:
correlation function of Abell clusters located in superclusters with  
at least 8 clusters (Einasto \etal 1997b).}
\includegraphics{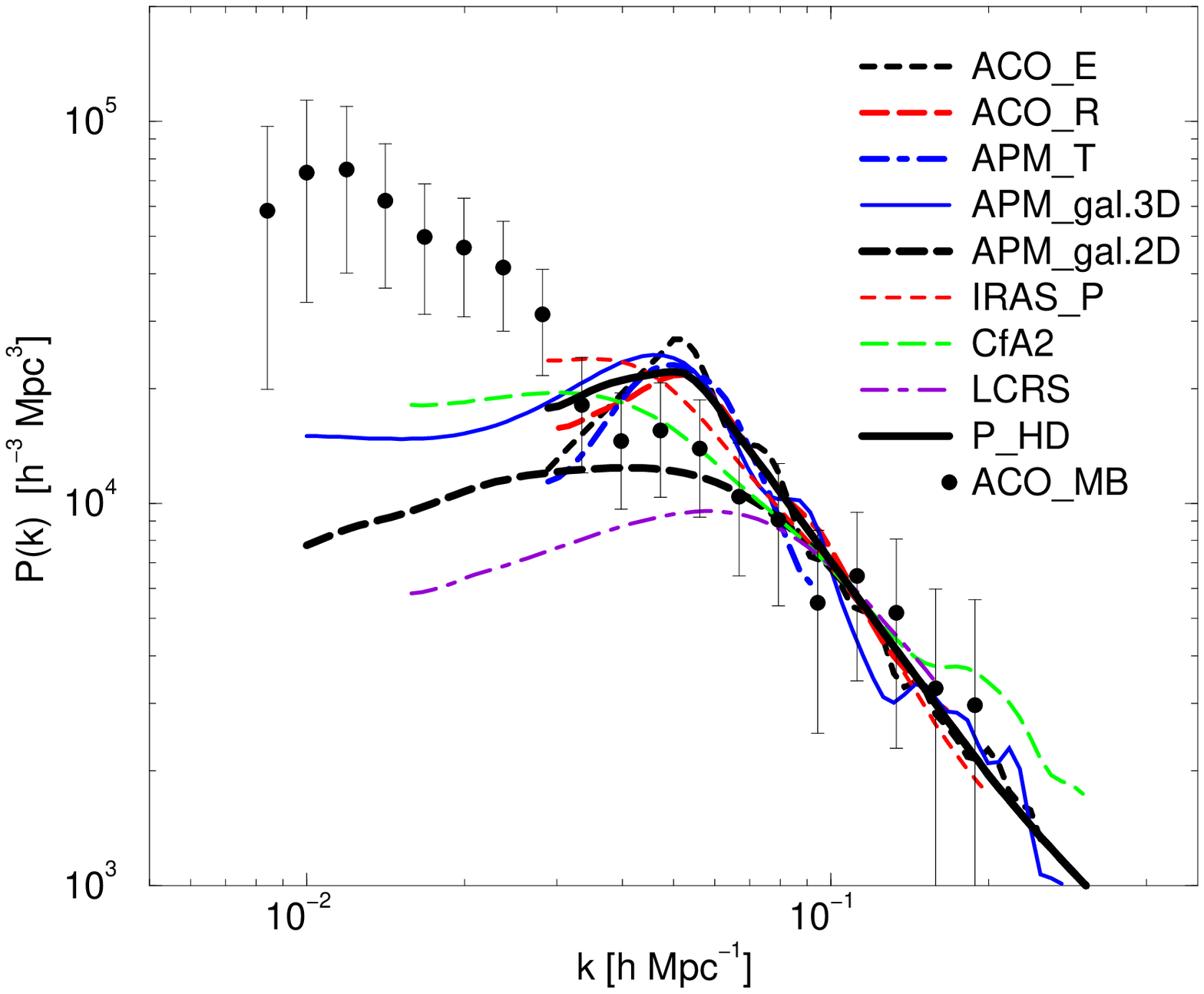}
\includegraphics{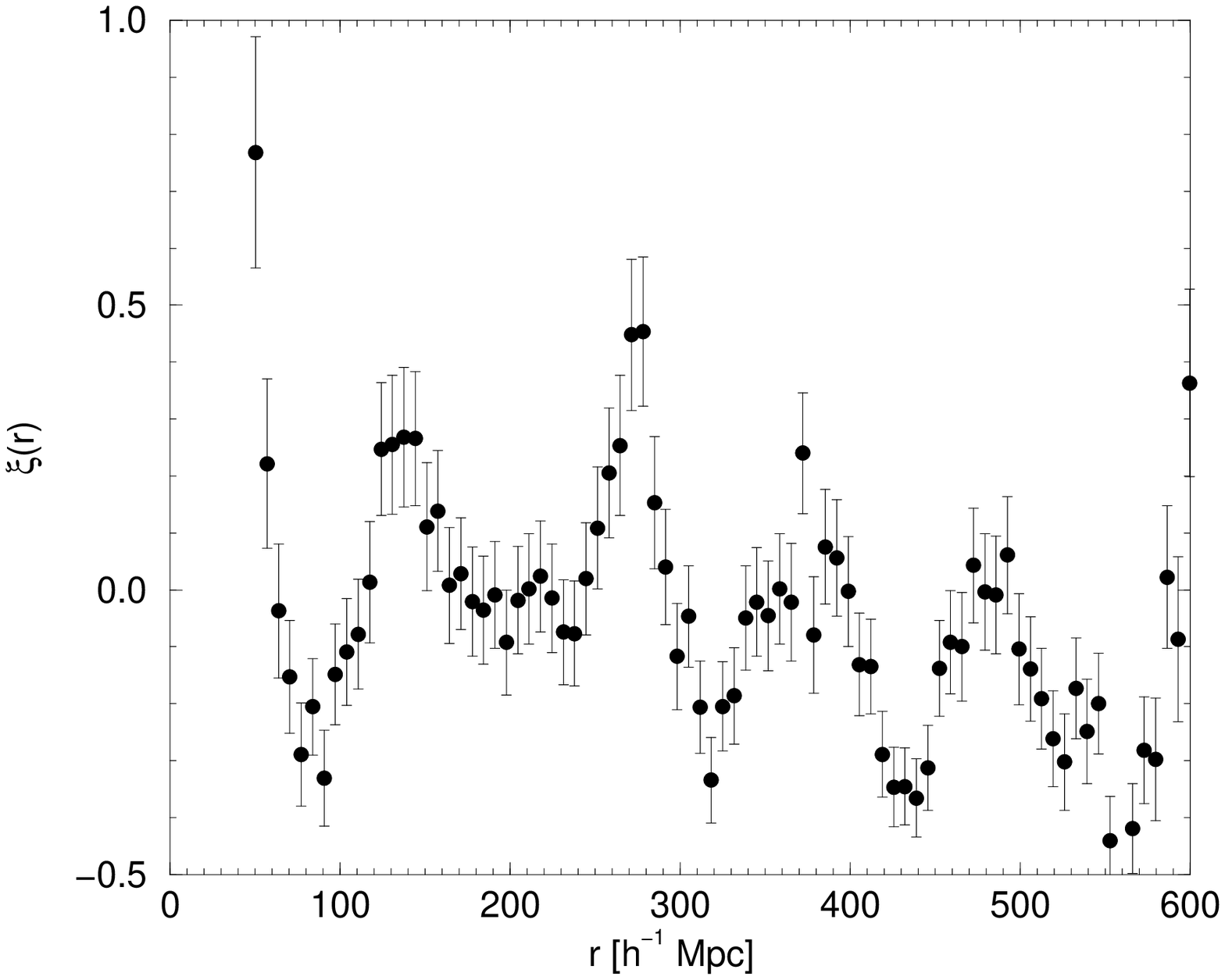}
\label{sp_obs}
\end{figure}

Power spectra of various galaxy and cluster samples are shown in the
left panel of Figure~\ref{sp_obs}; the correlation function of Abell
clusters in rich superclusters is plotted in the right panel. We see
that the correlation function of rich clusters is oscillating with a
period 120~\Mpc.  The mean power spectrum of clusters and deep galaxy
samples has a sharp maximum on the same wavelength.  These properties
are similar to properties of the regular rod model.

\begin{figure}[ht]
\vspace*{4.0cm}
\caption{Correlation functions of clusters in rich superclusters: the left
panel shows superclusters of richness $N_{cl} \geq 3$ of X-ray
selected clusters and active galaxies; the right panel shows superclusters
of APM clusters of richness $N_{cl} \geq 8$.
}
\includegraphics{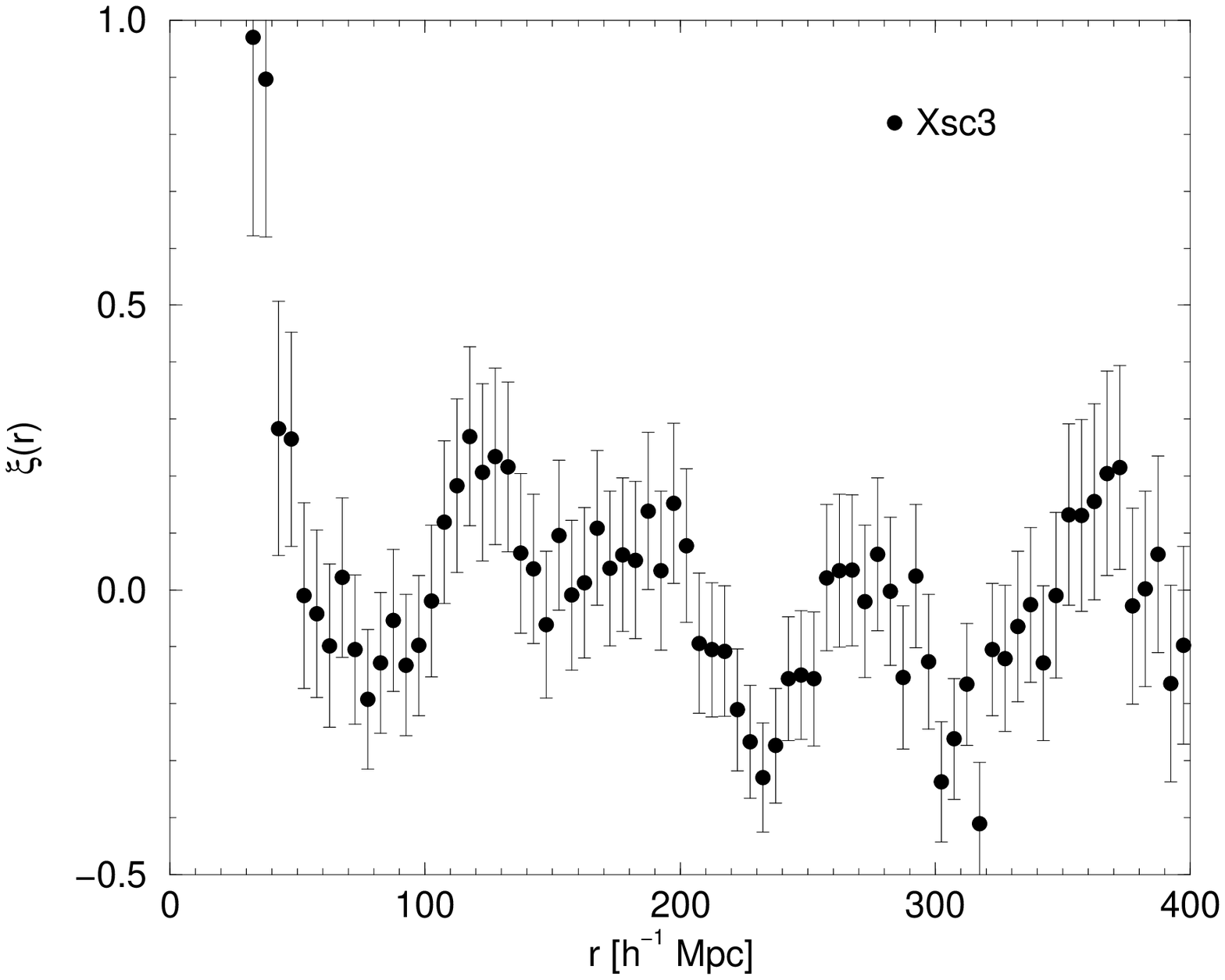}
\includegraphics{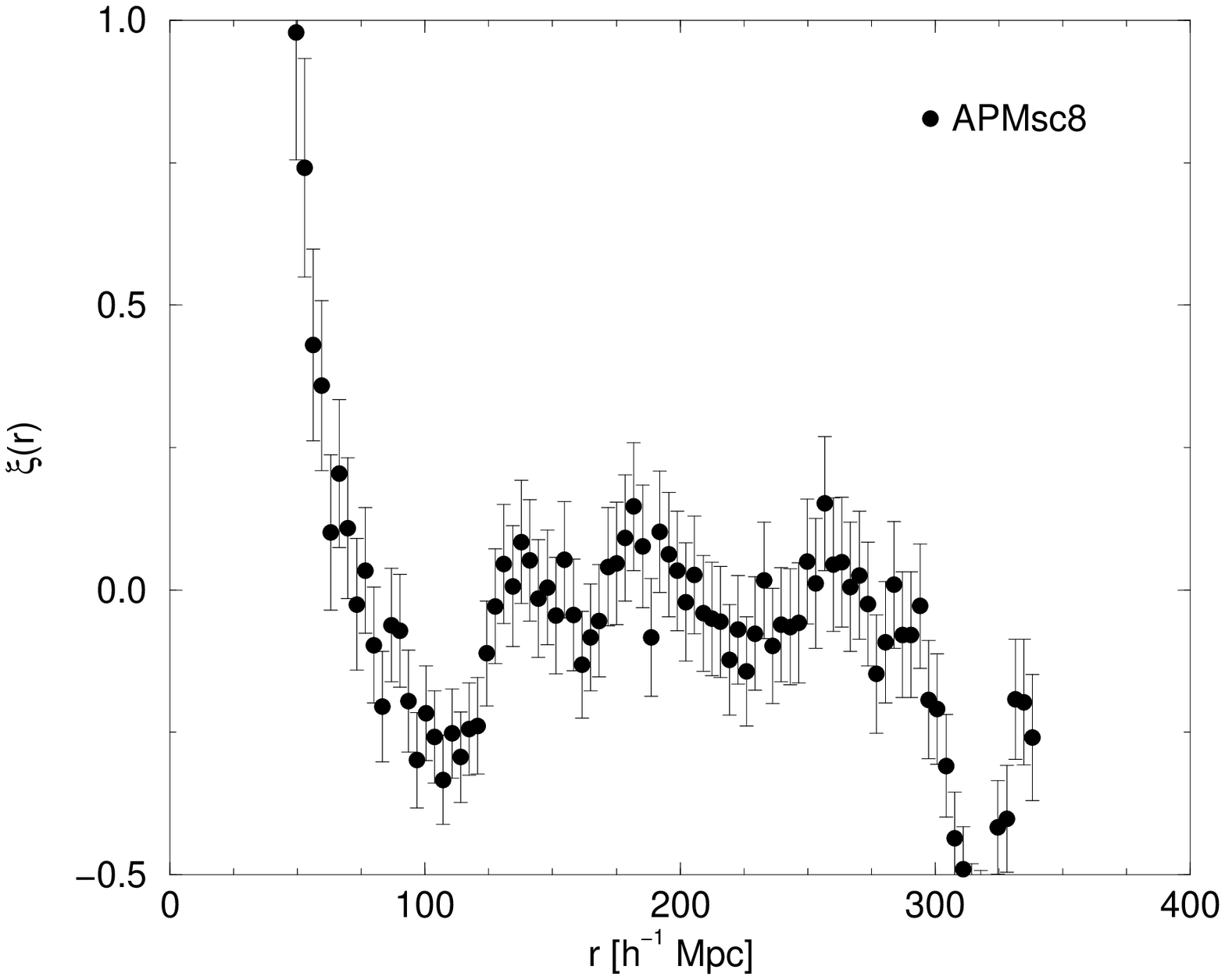}
\label{corr}
\end{figure}

\subsection{APM and X-ray Clusters}

The largest weight in the mean power spectrum plotted in
Figure~\ref{sp_obs} has the Abell cluster sample as it is the deepest
and the largest in volume.  However, Abell clusters were selected by
visual inspection of Palomar survey plates and can be influenced by
unknown selection effects.  Thus an independent check of the power
spectrum and correlation function is needed.  We have used A(utomated)
P(late) M(easuring) survey clusters and X-ray selected clusters and
active galaxies for comparison.

In Figure~\ref{clus} we show both optically and X-ray selected
clusters, demonstrating that rich superclusters are well seen in both
cluster samples. The richest Abell cluster superclusters are also rich
in X-ray clusters.  There exist differences in detail: not all Abell
clusters are strong X-ray emitters, and some X-ray luminous clusters
are not included in the Abell catalogue.  Similarly, APM clusters do
not coincide always with Abell clusters in the same volume. However,
the comparison of spatial distributions of cluster samples shows that
APM and X-ray selected clusters form rich superclusters very close to
rich superclusters of Abell clusters.

Correlation functions of APM and X-ray selected clusters in rich
superclusters are also rather similar to the correlation function of
Abell clusters in rich superclusters: compare Figures~\ref{sp_obs} and
\ref{corr}. Again, there are differences in details: the correlation
functions of APM and X-ray clusters have a secondary maximum at a
separation of $r \sim 190$~\Mpc\ which is absent in the correlation
function of the Abell cluster sample.  This feature is due to the fact
that the APM sample (and partly the X-ray selected sample) is
dominated by two very rich superclusters surrounding an exceptionally
large void between Horologium-Reticulum and Sculptor superclusters
(Einasto \etal 1997d), such that the maximum of the correlation
function is shifted.  On the other hand, the Abell sample contains a
dozen very rich superclusters and differences in void diameters cancel
each other out.  Due to the smaller volume covered, the APM cluster
sample is less suitable for large-scale studies.

\vskip0.3cm
The  observational evidence can be summarised as follows:
\vspace*{-2mm}
\begin{itemize}

\item{} Rich superclusters and voids form a quasi-regular lattice;
\vspace*{-2mm}
\item{} The mean diameter of cells (the mean separation of rich 
superclusters across voids) is $120 - 130$~\Mpc. 
\end{itemize}

\section{Theoretical Implications } 

\subsection{Evolutionary Toy Models }

To investigate the formation of a quasiregular cellular structure we
have performed numerical simulations of the evolution of a toy model
with two-power-law spectrum of index $n_l=1$ on large scales and $n_s
=-2$ on small scales, and a sharp transition at wavenumber
$k=0.05$~\hmpc\ (Frisch \etal 1995).  This power spectrum is rather
similar to the observed power spectrum.  The two-power-law model
reproduces well the distribution of rich clusters of galaxies.
Clusters form long filamentary superclusters and a quasiregular
supercluster-void network.  The correlation function and the power
spectrum of the two-power-law model approximates well the observed
functions.  However, this model is phenomenological. Physically
motivated models based on Dark Matter and inflation properties are
needed for comparison.

\subsection{Cold Dark Matter Models with a Bump}

The standard high-density CDM model ($\Omega_m=1$) has a power
spectrum very different from observations. On small scales the COBE   
normalised CDM model has an amplitude of the spectrum which is much higher
than the amplitude of the observed power spectrum of galaxies.  As
discussed above, this is difficult to explain.  LCDM models with
cosmological term have better agreement with the observed power
spectrum, but they lack the feature at the dominant observed scale.
The Low-density Mixed Dark Matter model fits better the observed power
spectrum,  but no feature is present either.

\begin{figure}[ht]
\vspace*{10.0cm}
\caption{ Upper left: power spectra of an LCDM model with and without
Starobinsky modification.  Upper right: power spectra of MDM models
with and without Chung modification.  Lower left: cluster mass
distributions for MDM models with and without Chung
modification. Observed cluster mass functions are shown according to
Bahcall \& Cen (1993) and Girardi \etal (1998). Lower right: angular
power spectra of tilted MDM models with and without Chung modification
(amplitude parameter $a=0.3$), compared with BOOMERANG (de Bernardis
\etal 2000) and MAXIMA I (Hanani \etal 2000) data. }
\includegraphics{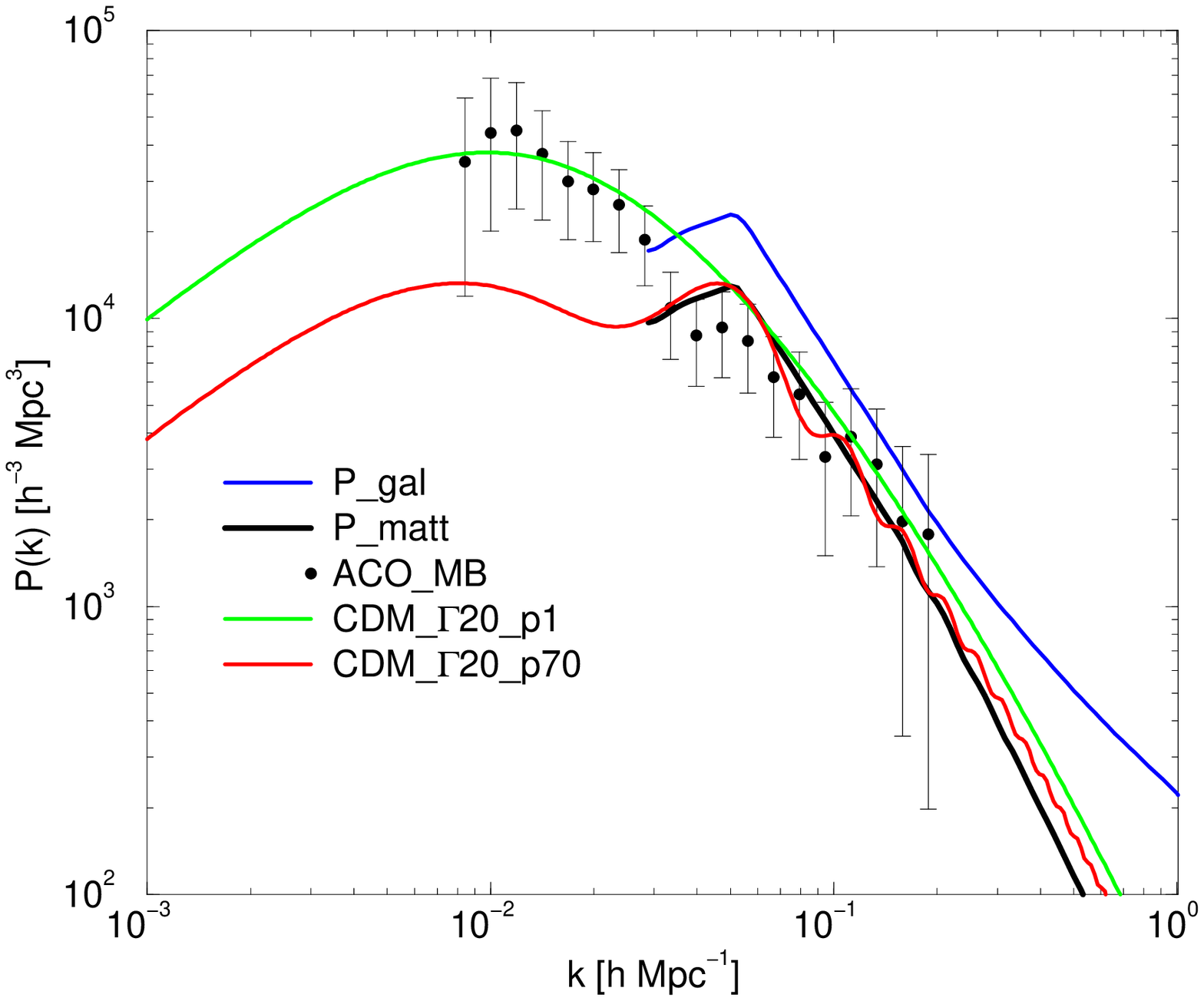}
\includegraphics{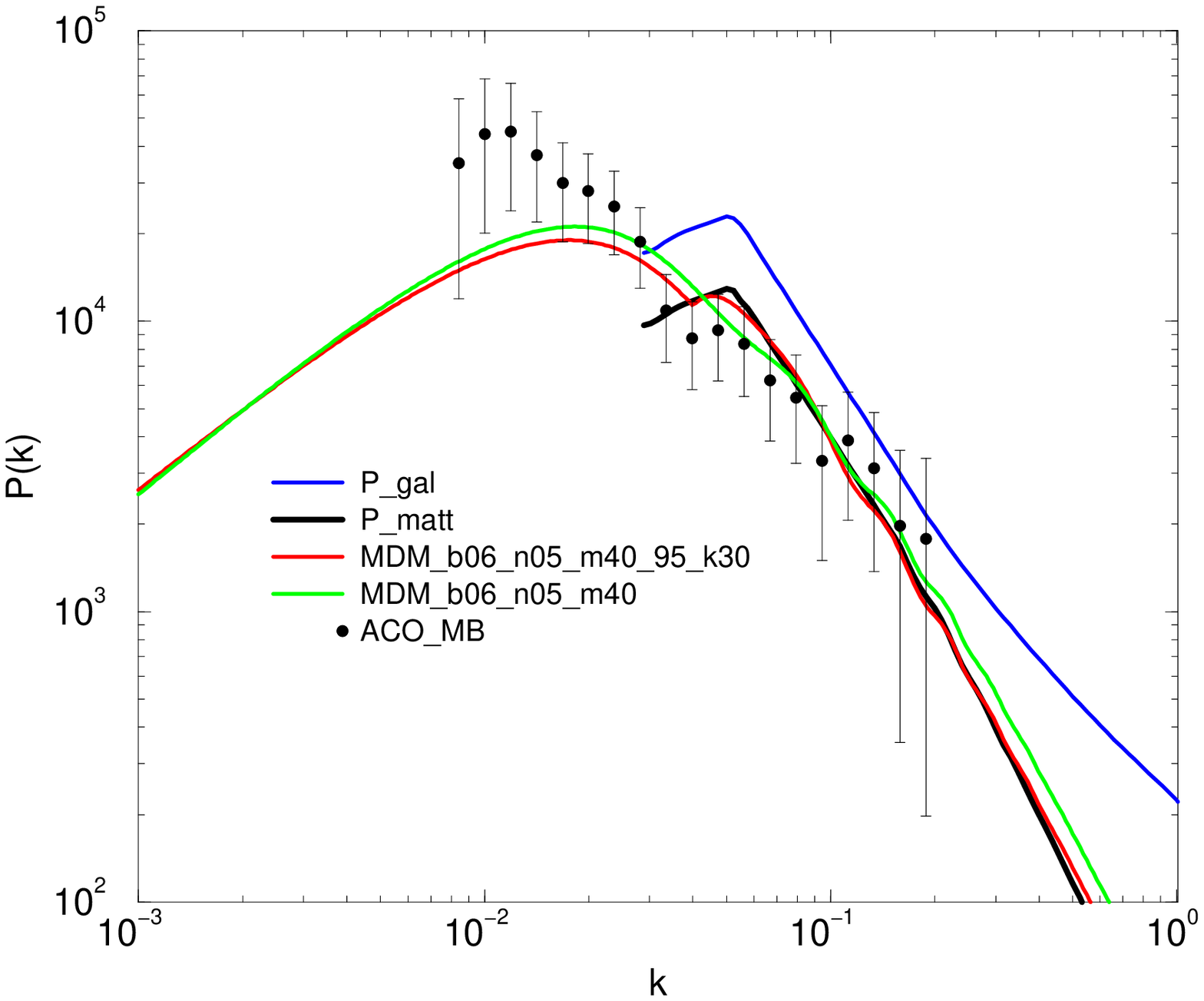}
\includegraphics{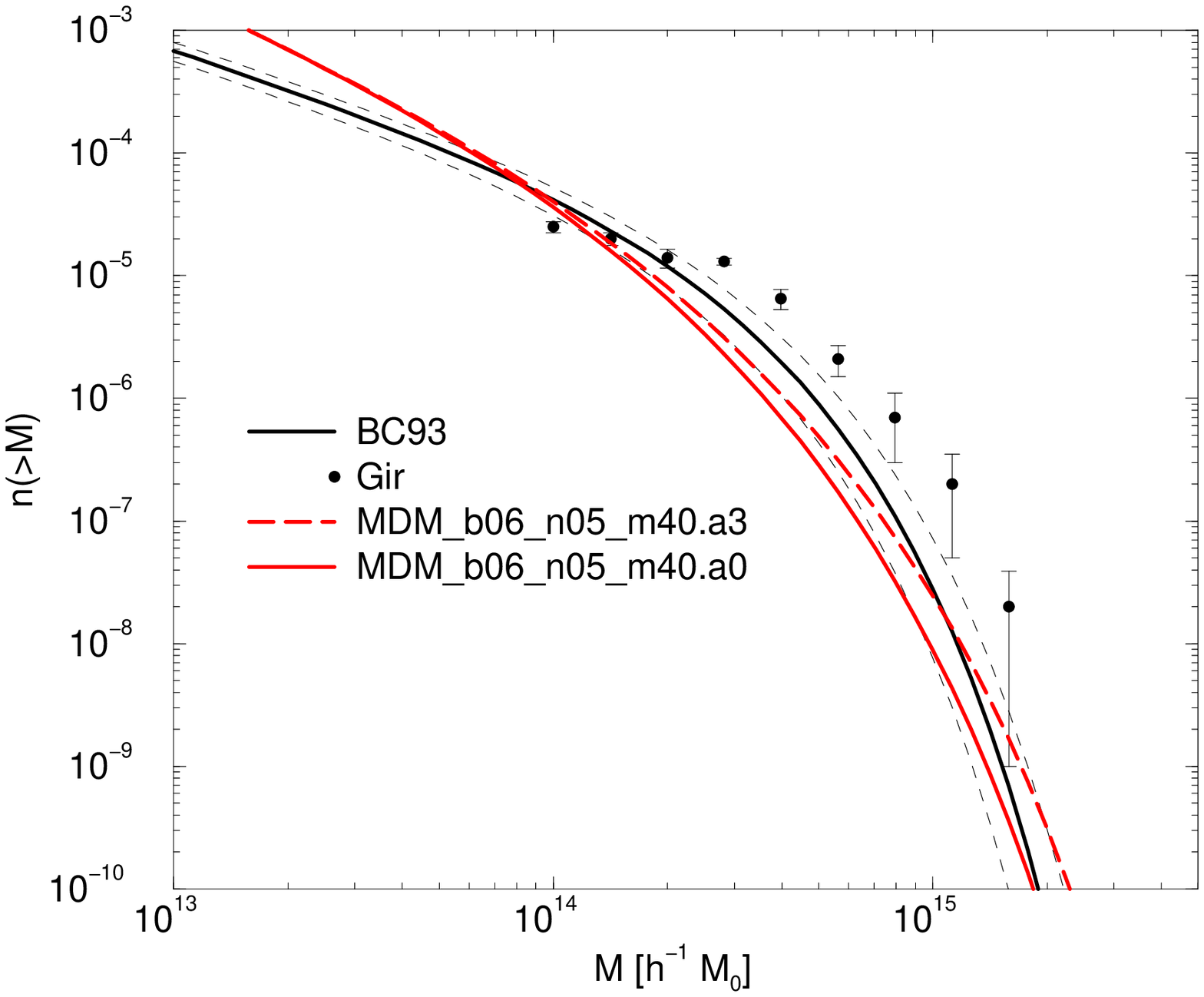}
\includegraphics{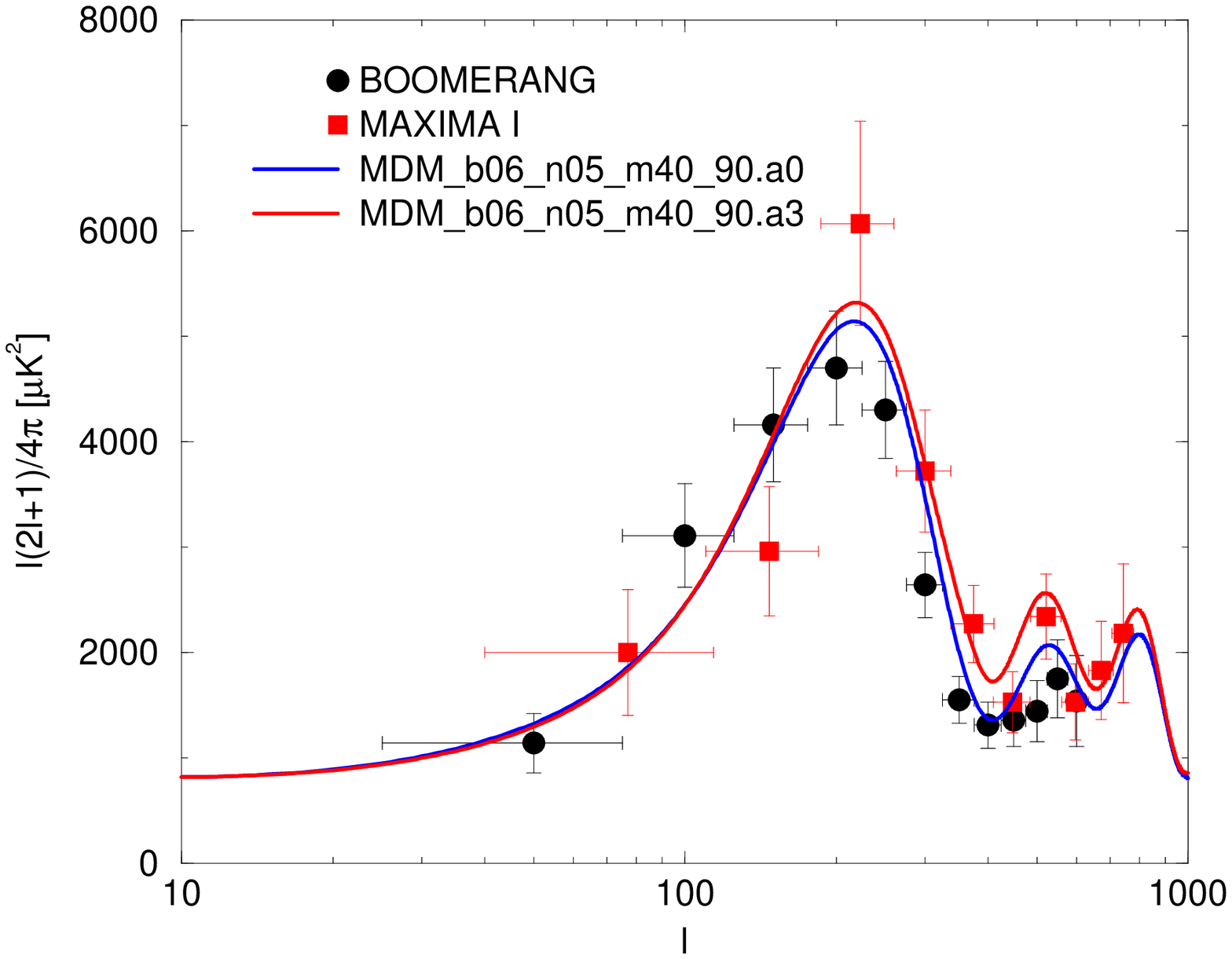}
\label{model2}
\end{figure}

In some variants of the inflation scenario primordial power spectra
have a bump or peak. The model suggested by Lesgourgues, Polarski \&
Starobinsky (1998) has a break in amplitude and peak; the model by
Chung \etal (1999) has a bump.  Parameters of the model are the
position and height of the bump (peak).  Power spectra of the
Starobinsky model are plotted in the upper left panel of
Figure~\ref{model2}.  Spectra are based on an LCDM model with shape
parameter $\Gamma= \Omega_m h = 0.2$, and bump parameters $p=1$ (no
bump) and $p=0.7$.  The upper right panel shows power spectra of an
MDM model with cosmological parameters: baryon density
$\Omega_b=0.06$, CDM density $\Omega_c=0.29$, HDM (neutrino) density
$\Omega_n=0.05$, vacuum energy density $\Omega_l=0.60$, and Hubble
parameter $h=0.65$.  This set of parameters is rather close to the
model suggested by Ostriker \& Steinhardt (1995) and confirmed by new
data by Burles \etal (1999), Riess \etal (1998), Perlmutter \etal
(1998), and Parodi \etal (2000) among others.  One spectrum
corresponds to the conventional MDM model, the other to the model with
Chung peak of amplitude $a=0.3$.  Calculations were made with the
CMBFAST package by Seljak \& Zaldarriaga (1996).
 
This Figure shows that a Starobinsky model with a bump is in good
agreement with earlier observational data on the galaxy and cluster
power spectrum but not with the new data by Miller \& Batuski
(2000). A mixed dark matter model with Chung peak of amplitude $a=0.3$
is in good agreement with all observational data.

Finally we have checked the Chung model with other independent data.
The lower left panel of Figure~\ref{model2} shows the cluster mass
function calculated using the Press \& Schechter (1974) algorithm.
These calculations show that the model with a bump produces more
clusters of high mass. However within observational errors of the
cluster mass functions of both models are in agreement with
observations.  A similar comparison with new CMB measurements is given
in the lower right panel of Figure~\ref{model2}. To improve the
agreement with CMB data we have used a tilted MDM model with spectral
index $n_l=0.90$ on large scales, keeping all other cosmological
parameters as in the model shown in the lower left panel.  These
calculations show that models with and without the Chung modification
are in satisfactory agreement with the CMB angular spectrum.  Thus we
conclude that this model represents well all available observational
data.

However, there are several problems not solved yet.  First, the
observational question: How large is the region of the regular
supercluster-void network?  Then the question concerning the
interpretation of models: How are spectral features (bump) and
regularity related?  We hope that this question can be answered with
numerical simulations of LCDM models with and without a bump in the
power spectrum, and such simulations are under way.  Finally, the
theoretical questions: Why is there a preferred scale and why does it
have a value of $\sim 130$~\Mpc?  What does the presence of a scale
and a regularity tell us on the inflation?

\section{Conclusions} 

\begin{itemize}
\vspace*{-2mm}
\item{} There exists good evidence on the presence of a scale of $\sim
130$~\Mpc: it measures the mean distance between rich superclusters
across voids.
\vspace*{-2mm}
\item{} There are variants of the inflation scenario suggesting the
presence of a characteristic scale (bump) in the power spectrum.
\vspace*{-2mm}
\item{}  Models with a bump in the power spectrum are in agreement with
other cosmological data using sets of parameters within acceptable
ranges. 

\vspace*{-2mm}
\item{} Cellular large-scale structure may be the end of the fractal
structure of the Universe, as suggested by Ruffini, Song \& Taraglio
(1988). 
\end{itemize} 

We should note that models neither predict the position of the bump,
nor the presence and extent of the regularity of the
supercluster-void network. We should remember the maxim by Eddington:
{\em no experimental result can be taken seriously if not explained
theoretically}.  Thus larger surveys are needed to clarify properties
of the supercluster-void network.  Also a further theoretical analysis
are needed to find the physical origin of the scale and the extent of
the regularity.

\section*{Acknowledgments}
I thank Maret Einasto, Mirt Gramann, Pekka Heinam\"aki, Volker
M\"uller, Enn Saar, Aleksei Starobinsky, Ivan Suhhonenko and Erik Tago
for fruitful collaboration and permission to use joint results in this
talk. Heinz Andernach kindly improved the presentation.  This study
was supported by the Estonian Science Foundation grant 2625.

\end{document}